# Running a distributed virtual observatory: US Virtual Astronomical Observatory operations

Thomas A. McGlynn[*a], Robert J. Hanisch[b,c], G. Bruce Berriman[d], Aniruddha R. Thakar[e]

[a]High Energy Astrophysics Science Archive Research Center, Code 660.1, NASA/Goddard Space Flight Center, Greenbelt, MD USA 20771; [b]U.S. Virtual Astronomical Observatory, 1400 16th Street NW, Suite 730, Washington, DC USA 20036; [c]Space Telescope Science Institute, 3700 San Martin Drive, Baltimore, MD USA 21227; [d]Infrared Processing and Analysis Center, California Institute of Technology, Mail Stop 100-22, 770 South Wilson Avenue, Pasadena, CA USA 91125; [e]Center for Astrophysical Sciences, The Johns Hopkins University, 3701 San Martin Drive, Baltimore MD USA 21218.

## ABSTRACT

Operation of the US Virtual Astronomical Observatory shares some issues with modern physical observatories, e.g., intimidating data volumes and rapid technological change, and must also address unique concerns like the lack of direct control of the underlying and scattered data resources, and the distributed nature of the observatory itself. In this paper we discuss how the VAO has addressed these challenges to provide the astronomical community with a coherent set of science-enabling tools and services. The distributed nature of our virtual observatory—with data and personnel spanning geographic, institutional and regime boundaries—is simultaneously a major operational headache and the primary science motivation for the VAO. Most astronomy today uses data from many resources. Facilitation of matching heterogeneous datasets is a fundamental reason for the virtual observatory. Key aspects of our approach include continuous monitoring and validation of VAO and VO services and the datasets provided by the community, monitoring of user requests to optimize access, caching for large datasets, and providing distributed storage services that allow user to collect results near large data repositories. Some elements are now fully implemented, while others are planned for subsequent years. The distributed nature of the VAO requires careful attention to what can be a straightforward operation at a conventional observatory, e.g., the organization of the web site or the collection and combined analysis of logs. Many of these strategies use and extend protocols developed by the international virtual observatory community. Our long-term challenge is working with the underlying data providers to ensure high quality implementation of VO data access protocols (new and better 'telescopes'), assisting astronomical developers to build robust integrating tools (new 'instruments'), and coordinating with the research community to maximize the science enabled.

## 1. INTRODUCTION

The United States Virtual Astronomical Observatory (VAO) is jointly supported by the US National Science Foundation (NSF) and the National Aeronautics and Space Administration (NASA) to support research using the myriad archival data resources available in astronomy. The VAO is itself distributed with nine major institutions. This paper discusses how the VAO operates to provide effective services to the community. It does not address the internal management of the VAO itself (see [1]), nor the internal processes used in software development except as they directly affect our science users (though see [2] for a review of the processes involved in making an astronomical data collection VO-enabled).

### 1.1 The environment of the Virtual Astronomical Observatory

The Virtual Astronomical Observatory was started in May 2010 as a five-year joint initiative of the NSF and NASA. The VAO is a collaboration whose major partners include the Space Telescope Science Institute (STScI), High Energy

---

[*]thomas.a.mcglynn@nasa.gov; phone 1 301-286-7743

Astrophysics Science Archive Research Center (HEASARC) at NASA's Goddard Space Flight Center, Johns Hopkins University (JHU), the National Center for Supercomputing Applications (NCSA), the National Radio Astronomy Observatory (NRAO), the National Optical Astronomy Observatories (NOAO), The Smithsonian Astrophysical Observatory (including the Astrophysics Data System, ADS), the NASA Extragalactic Database (NED) and the Infrared Science Archive (IRSA) both at the Infrared Processing and Analysis Center (IPAC), and the California Institute of Technology. Associated Universities, Inc. (AUI) provides business management services.

The goal of the VAO is to facilitate astronomical research using distributed data resources in the community. The VAO uses resources already available and makes it easier to add new resources into the astronomical virtual observatory. The VAO is successor to the National Virtual Observatory (NVO) and continues its role in helping to define and develop the standards and infrastructure through which science services are to be delivered. The VAO acts the US representative in the International Virtual Observatory Alliance (IVOA), a consortium of the national astronomy virtual observatory efforts. The IVOA acts as the clearinghouse through which virtual observatory standards are developed and approved. Within this paper we use the acronym VO to mean the worldwide efforts in astronomy to share data, and VAO to refer specifically to the US VAO program.

During its first year, the VAO has developed several powerful software capabilities that are now available to the community. These include a data discovery tool which is an effective browser for the entire data holdings of the virtual observatory, a tool to retrieve and analyze spectral energy distributions, a tool for large scale cross-correlations, and tools for the analysis of time series data. To make effective, reliable, and consistent science tools in our distributed environment using distributed data has been a challenging task. This paper describes our general operational approach to ensuring that our science services are useful and robust.

## 1.2 Data in the VO

Among the sciences astronomy has been one of the leaders in the development and utilization of public archives. There are many factors that have led to this. During the 1980's and 1990's FITS was almost universally adopted as a data standard not just for interchange but also for archival storage. The astronomy data community is relatively small, but has many different data providers. Much of astronomy is focused on discrete and easily distinguished objects: stars, galaxies and the like. Since these objects have fixed positions (unlike, say, solar events, or terrestrial weather patterns) serendipitous observations can easily be identified and extracted. The use of CCD's as detectors has provided a uniform and commonly understood data structure in key wavebands. The need to view objects in multiple wavelengths (and over different time periods) leads astronomers to look to sources from multiple ground and space based observatories. With relatively limited non-scientific applications, there are few incentives to restrict access to astronomical data for commercial advantage.

By the turn of millennium, many observatories found more articles were being published from archival research than from the investigations for which observations had originally been made. Major new initiatives, such as the Sloan Digital Sky Survey, made providing a general public record to be used by astronomers world-wide a primary objective of the project from inception.

Over the past decade, the IVOA has adopted a series of data standards that provide for the distribution of catalog, imaging and spectral data and general archival resources [3]. A standard index for these resources is available in the VO *registry*. The registry is a distributed resource—duplicated by multiple VO organizations and accessible through a standardized protocol. A few numbers show the scope of the VO: as of June 2012 there are more than 10,000 distinct resources indexed in the registry from over 70 different institutions.

A key aspect of the VO is that each site is autonomous. Data providers are responsible for their own data and metadata. Data providers are responsible for adding their own resources to the registry and for the accuracy of that record. Not surprisingly both data and metadata are of uneven quality.

## 1.3 The operations challenges

When building science services in this environment the VAO faces several key challenges:

1. We must provide reliable and high quality services given uneven quality of the available datasets. This area has been our primary focus the first year of VAO operations.

2. Data services must work effectively within the VAO as integrated suite despite our inherently distributed nature.

3. We must enable handling of large quantities of data. User tables often exceed a million rows and we wish to correlate against system tables which now have billions.

Different challenges come to the fore in different science activities. For example, understanding the quality of services is central to our data discovery tool, while data volume is an issue in doing massive cross-correlations.

## 2. ENSURING QUALITY OF SERVICE

The VAO works in a complex ecosystem of data providers and software agents. To work effectively in this world requires understanding which resources are available for use, and the extent to which they can be trusted. A significant fraction of our VAO operations effort is spent in monitoring and validation of all VO resources. Of course we monitor our own internal capabilities to make sure that VAO supported services are running. However if one of our services fails not because of anything we directly control, but because the archive providing a critical dataset has gone down, our users only see that the system is broken.

### 2.1 Internal monitoring

VAO services are currently provided from all of the constituent institutions. To ensure that all VAO tools are operational, including both services we provide to the community and the capabilities that support VAO's internal activities, all services are monitored hourly. A database of results is maintained. When a problem is detected, a second check is run a few minutes later to avoid false alarms due to network glitches. If the second request also fails an automated E-mail is sent to the responsible party for that service, to the VAO operations manager, and to any other VAO personnel interested in monitoring that site. Monitoring is put in place before services are released to the public. Figure shows the average uptime of VAO services during the past year. While the averages are generally well in excess of 95% we had hoped to achieve consistent uptimes of greater than 99%. Budget cutbacks have derailed our original plans for providing backup capabilities for all science services, but while we have had several significant interruptions of our internal services (including the monitors themselves), few of these have affected the science capabilities provided to our users.

The chart shows several incidents with significant downtime, but in most cases the effects on science services to the community were limited, e.g., an outage at the NCSA affected our validation services but not the science services then available.

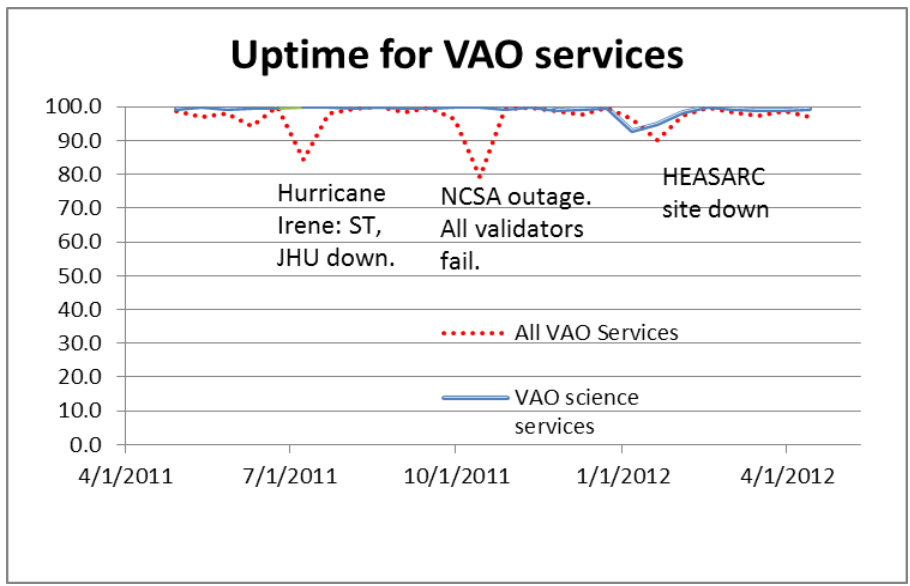

Figure 1. Percentage uptime for VAO services.

## 2.2 External monitoring

The VAO is not a primary data provider in the VO. We provide services that use resources provided by other institutions. To ensure that these resources are available the VAO tests all VO data-providing sites once per hour. Since some sites provide many services, e.g., the Vizier site provides access to thousands of tables, it is inappropriate to test every service every hour. Rather we test one service of each of the major types of services supported at the site each hour. This test is intended to be quick with a limited regression against a known result. It does not do any independent check that the result is correct or correctly follows accepted protocols. Recently we have begun recording all monitoring results in a database so that we can create statistics on the up and down times for given services.

When a service fails, the VAO operations monitor informs the responsible party for the site and works to ensure that the service is brought back up as soon as possible. If an extended down-time is anticipated, a notice is made available at the VAO notifications page to let all interested parties know of the problem. While such notices are publically visible today, we plan on adding notices to the VAO home page to make sure that issues get high visibility.

Typically several issues are discovered each week. The issues are logged in our internal issue tracking system. Between December 2010 and June 2012, 290 service interruption tickets were defined. Often the site is already aware of the problem before we notify them, but many times our message is what lets them know of the problem. Given the lack of any lines of authority between VO institutions, this collaborative effort to ensure that downtimes for services are minimized is essential and the continual monitoring by the VAO has a clear positive effect in ensuring uptime of the VO's distributed services.

Occasionally services are abandoned by their creators and simply no longer work. Without any action these remain in the VO registry and can mislead users who anticipate getting results from them. Automated searches are delayed while connections time out. Every few months the VAO reviews services that are failing without giving any useful results. After a review, in which the responsible parties for the service are informed if there is any valid contact information, such obsolete services are deprecated. They are not fully removed from the registry, but they are marked as non-functional and are not returned in our default queries nor used in our discovery services.

## 2.3 Service validation

The VAO validates all registered VO data services against the VO standards. This includes all services implementing the four currently approved data retrieval capabilities: Simple Cone Search, Simple Image Access, Simple Spectral Access and Table Access. In addition VO registries' conformance to the Open Archive Interface (OAI) protocol is checked. A set of validators, some developed at the VAO (Cone Search and Simple Image Access) and others developed in Europe (Simple Spectral Access at the University of Paris [4] and Table Access using Mark Taylor's TAPlint [5]), are run against all known services and a record of all errors found is recorded in a database. Each registered services is validated roughly once per month. The validators send a series of requests to the service and attempt to rigorously test each service for compliance to the standard. Currently we treat a service that fails any requirement of the standard as failed, but we anticipate transitioning to a graded system where the fractional compliance to the standard is assessed on a per service basis. Results of our validation for cone and simple images access services are shown in Figure 2.

This chart distinguishes between *all* VO services, and services that are hosted at institutions that are collaborators within the VAO. It is clear that strict compliance with standards is a challenging goal that almost half of registered serivces fail to meet. On the one hand it should be recognized that many of the failures are in meeting aspects of the standards that do not really affect the science quality of most results, e.g., the details of error formats, or providing single precision floating point results where the standard mandates double precision. On the other hand this validation is purely of the technical implementation of the services syntax with no evaluation of actual data quality. Our hope to provide some level of science review of the quality of the services was a victim of budget cuts to the VAO.

A sharp dip in compliance occurred in March 2011 when Vizier, a major data provider implemented a protocol breaking change to their services. We informed Vizier of the problem which they immediately corrected, but since we only retest each service once a month it took time for the repair to show up in the statistics.

Services provided by VAO host institutions have a significantly better record of passing the validation tests. This difference is ascribable to our attention to, and frequent internal discussion of, validation results.

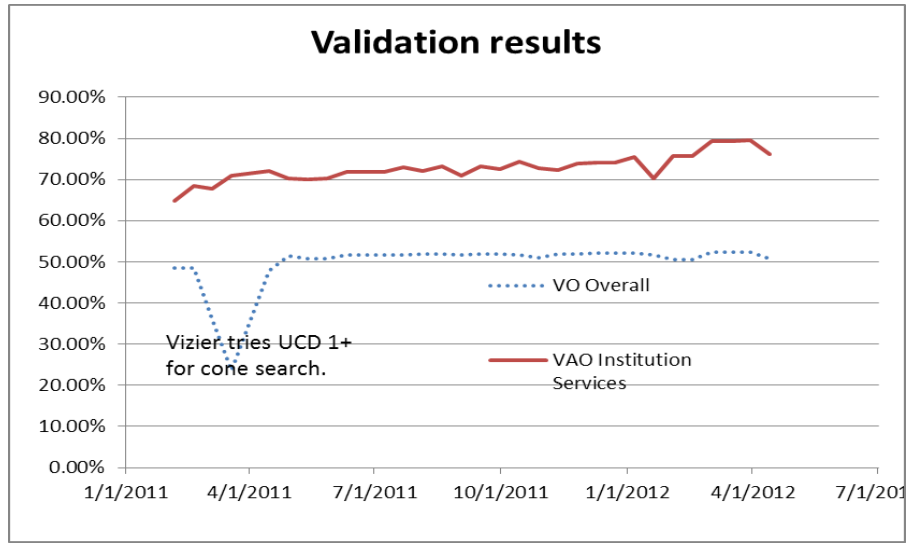

Figure 2. Fraction of services passing automated validation.

We periodically create summaries of validation issues for each institution, including suggestions for how validation problems could be resolved. These are sent to all responsible hosts and we have noted a significant increase in compliance after we have sent out such reports. Although our detailed database goes back only to early 2011, our validation activities go back several years. Initially the total pass rate for services was less than 5%.

### 2.4 Interruption notification

Monitoring and validating services gives us a mechanism for detecting problems. The VAO notification service enables us to tell users about problems detected by these and also to inform users of expected downtimes due to system maintenance at one of the VAO sites. Currently the notification service is provided as a separate web page, but a summary of the system status will soon be provided directly on the VAO home page with links to a detailed status of all services.

### 2.5 Centralized logging

With services running at multiple institutions, a centralized logging facility is useful in several areas. Low levels of activity can show a problem with a service. It also allows the calculation of usage metrics. A centralized facility helps us to understand the relationship between usage of our different services. With only a few months of logs since our initial release of many of our science services, the analysis of these logs has so far been limited to tracking the growth of the usage of our services. So far we have been logging the web hits associated with VAO sites and services, we have not yet begun to log details of service requests (e.g. SQL queries, image and spectrum access, etc.). Detailed analysis of logs has been deferred due to budgetary considerations.

## 3. MAKING THE VAO A COHERENT ORGANIZATION

Our users generally have no interest in the internal makeup of the VAO and its constituent organizations. For the VAO to succeed its operating capabilities need to work together seamlessly. VAO operations use standard web and VO technologies not to conceal the distributed nature of our organization, but to make it irrelevant from the user's perspective.

### 3.1 The VAO domain

The VAO owns the usvao.org domain. While the home page is hosted at Caltech, subdomains are also found at NOAO, STScI and NCSA. We anticipate additional USVAO domains at the other VAO host institutions as their services become more visible to our community. All VAO services will be provided through USVAO web addresses with different sub-hosts at different institutions. While we have experimented with a tighter integration of our system using

Apache's proxying from a single central host, initial results suggested that this coupled the file systems of the distributed sites too tightly. While it is straightforward to organize the primary web pages using proxies, the myriad included URLs (icons, logos, headers, footers and JavaScript) become easily confused. Using proxies may be more appropriate in later stages of the VAO when capabilities are changing less rapidly.

### 3.2 The VAO website

Recently the VAO moved to using WordPress as a content management system (CMS) for its home web site. WordPress, along with its myriad plugins, provides excellent functionality with no financial risk and a short learning curve. Limited studies of other content management systems were conducted. Some, like Drupal, seemed more powerful but it was unclear that we needed the additional capabilities and they exacted a cost of a significantly more difficult startup.

Our content management system is intended to be used for 'standard' pages: news, documentation, help and the like. Our science services can have very complex requirements, e.g., the data discovery tool uses complex AJAX-based pages. By separating out these very specialized web pages, it becomes much easier to support our web presence using a relatively simple CMS.

A key advantage of the Web-based CMS approach is that VAO developers at multiple institutions can more easily add and update to our Web site. Previously our web site had required accounts on the servers hosting the web pages.

For pages hosted at sites other than our primary web server, we provide a set of templates that ensure a consistent look and feel. We get the advantages of a CMS for the bulk of the site that has relatively straightforward pages, but as much flexibility as necessary when complex pages are being put together.

### 3.3 Communication between VAO tools

With the release of several VAO science services this January, we are beginning to face the issue of transport of data between services. For example, our data portal is hosted at STScI, our cross-correlation service at IPAC and our SED analysis tool runs on the user's machine. Results need to be able to move from one site to another. We have only begun to address this issue but two VO standards are particularly applicable. The Simple Application Messaging Protocol (SAMP) [6] allows for data transport between services a user is running, including between web-based and local user services. For the relatively small (of order a few megabyte) tables that we have been passing in our initial applications, SAMP provides adequate support.

However, particularly as we begin to do very large correlations, we anticipate providing VOSpace support at our sites [7]. VOSpace is essentially a VO specific approach to cloud storage. Users will be able to persistently store large amounts of data near the analysis tools that are to use them. Deployment of VOSpace has been delayed until after we developed the services that will use them. It is a priority in our current development.

Another approach used in our cross-correlation service is to try to get as many of the large datasets that a user is likely to want to use co-located at a common, central site. The cross-correlation service's site hosts a copy of the popular giga-row catalogs that are typically used in large scale astronomical cross-matches. Replication of large resources will play a significant role in VAO operations as we scale to support very large queries. With the cost of storage dropping much more quickly than bandwidth between sites increases, it becomes increasingly cost effective to provide multi-point access to key resources. We are *not* advocating a "data center solution," since at any given time a significant fraction of the VO data will be outside of data centers – the VO is unavoidably a distributed environment.

The VAO has only begun to face the operational implications of the very large data volumes that we anticipate. Further development of VO standards is essential here. Standard interfaces for subsetting large datasets, such as cutouts of images and filters on data tables, are needed and some development is underway. It is not practical to download a gigabyte image to extract a 100x100 pixel cutout.

While most VAO services currently require no special access privileges, it is clear that support for power users of the kinds that will generate large data flows will likely involve restricting access through accounts with user authorization and authentication. Already some VO tools can trigger what are essentially inadvertent denial-of-service attacks against data providers. The VAO has in place a single sign-on facility where a user can sign on once and their authorization is automatically forwarded to new services as needed.

# 4. LESSONS AND FUTURE PLANS

The VAO has been operational for just over a year. VAO-developed science services have been available to the public since January 2012. Several themes have become clear in our operations. Monitoring and validation of both our science tools and the data services they use is essential in the very distributed environment of the VO. A combination of automated checks and manual follow up is essential in the collaborative VO environment. Measurement of how well VO developers are able to implement standards is crucial feedback to the future development of the VO infrastructure.

For the VAO to be effective, its many interacting components need to be alive, operating correctly, responding promptly, and giving scientifically relevant responses. As problems arise—and they surely will in a distributed system—users need to be able to easily determine what is working and what is not, and system maintainers and operations staff need adequate and timely information about faults in order to respond and repair problems promptly.

The centrifugal pull of our distributed VAO organization needs to be continually addressed by working to ensure consistency of the interfaces and centralization of capabilities like logging. Our adoption of a CMS to manage the bulk of our web site helps us here but to make this feasible we needed to recognize that not every page in our web site needs to be within the CMS system.

A key question that we are still working to address is how to handle the very large data volumes that will flow through the elements of a successful VAO and VO. With strategic replication of services we can lessen these volumes but this only ameliorates the problem, it does not solve it. Implementation of VO protocols like VOSpace is essential and ongoing, and further standards are needed here.

Our biggest challenge is to meet the growing expectations of the astronomical community—especially as data volumes explode over the next few years—within our available resources.

## ACKNOWLEDGMENTS

This paper describes work done with the support of the US Virtual Astronomical Observatory. The VAO is jointly funded by the National Science Foundation (under Cooperative Agreement AST-0834235) and by the National Aeronautics and Space Administration. The VAO is managed by the VAO, LLC, a non-profit 501(c)(3) organization registered in the District of Columbia and a collaborative effort of the Association of Universities for Research in Astronomy (AURA) and the Associated Universities, Inc. (AUI).